\documentclass{emulateapj}

\usepackage{natbib}

\shorttitle{The 10~$\mu$m infrared band of silicate dust}
\shortauthors{Tamanai et al.}
\begin{document}
%
%
\title{The 10~$\mu$m infrared band of silicate dust: \\
A laboratory study comparing the aerosol and KBr pellet techniques}
%
%
\author{A. Tamanai and H. Mutschke}
\affil{Astrophysical Institute and University Observatory,
Friedrich-Schiller-University Jena, Schillerg\"asschen 3, D-07745 Jena, Germany}
\email{akemi@astro.uni-jena.de}
\author{J. Blum}
\affil{Institut f\"ur Geophysik und Extraterrestrische Physik,
Technische Universit\"at Braunschweig, Mendelssohnstrasse 3,
D-38106 Braunschweig, Germany}
\and
\author{G. Meeus}
\affil{Astrophysikalisches Institut Potsdam,
An der Sternwarte 16, D-14482 Potsdam, Germany}
\begin{abstract}
The profile of the silicate 10~$\mu$m IR band contains important information about the evolutional stage 
of dust in circumstellar environments and the possible ongoing process of planetesimal formation. In order to extract this 
information, the observed band profiles are compared with calculated or laboratory-measured 
absorption cross sections of amorphous and crystalline grains with different sizes and compositions. 
We present in this study the first laboratory measurements of the 10~$\mu$m band profiles of nonembedded, 
i.e. free-flying, particles of amorphous and crystalline Mg$_2$SiO$_4$ (with two different particle shapes), 
amorphous and crystalline MgSiO$_3$, and crystalline olivine. 
We compare the spectra with those measured on embedded grains and discuss the potential of the new 
experimental method for comparison with observed spectra, as well as for future studies of 
agglomeration and surface manipulation of the grains. 
\end{abstract}



\section{Introduction}
Dust particles are important players in astrophysical processes such as star and planet formation. Changes in size, composition, and crystallinity of the dust grains indicate dynamical processes (e.g.~in accretion disks) and hold a key to understanding specifically the early formation of terrestrial (Earth-like) planets in the environments of young stars. The most important technique, which makes it possible to obtain information about the dust populations in different objects (and nowadays even at different regions within one object), is infrared spectroscopy. The most important role here is played by the 10~$\mu$m Si-O stretching band of silicate dust because silicates are the major dust component in most dusty media in space, and this band can be observed from the ground. 

In fact, solid particles in the interstellar medium consist mainly of amorphous silicate dust grains \citep[see the review by][]{whi92}, whereas presence of crystalline silicates has been confirmed via mid-infrared observations in circumstellar disks around the relatively bright young Herbig Ae/Be stars \citep[e.g.][]{mal98,mee01}, for evolved stars \citep{mol02}, the T Tauri stars \citep[TTSs;][]{mee03}, the TTS Hen 3-600 \citep{hon03}, and the Vega-type stars HD 145263 and $\beta$-Pic \citep[e.g.][]{hon04,oka04}. Analysis of the infrared emission bands and especially the 10~$\mu$m band profile of such silicate grains has allowed one to find indications of grain growth and crystallization that trace the early evolution of young circumstellar disks toward planetary systems \citep{bo05}. However, the analysis of observational infrared spectra is very complex for an accurate interpretation, since not only mineralogical properties such as crystallinity and chemical composition determine the shape and intensity of the distinctive spectral feature, but also size, shape, and agglomeration of the grains exert an influence. Thus, exhaustive laboratory studies of these influences on the silicate dust bands are essential for an appropriate interpretation of the observed spectra. 

Currently, band profiles that are used for comparison with observed spectra are mostly calculated, assuming simple geometrical models such as spherical or ellipsoidal grain shapes. The predictions of these calculations are uncertain because in reality the grain shape might be irregular. So far, laboratory measurements on real particle ensembles were not able to provide exact band profiles for a direct comparison due to the use of an embedding medium (KBr) in the sample preparation, which changes the band profile substantially by the influence of its electromagnetic polarization \citep{fab01}. In this Letter, we present a new method for laboratory measurements of extinction spectra avoiding the influence of electromagnetic interaction with solid embedding media. Applying the aerosol technique \citep{hin99}, we obtain for the first time extinction spectra in the mid-infrared region of both amorphous and crystalline silicate grains \citep[for amorphous SiO$_2$ particles, see][]{tam06}, which are directly comparable to observed spectra. We investigate the differences to KBr-measured band profiles by deriving peak positions from the band profiles measured with both methods. Furthermore, we use electron-microscopic investigation of the grains to associate morphological (size, shape, agglomeration) influences on the measured profile.
\section {Experimental procedures and samples}
Regarding the aerosol measurements, we used a White-type long-path infrared cell (MARS-8L/20V, Gemini Scientific Instr.) of about 20~m path length, attached to a Fourier transformation infrared spectrometer (FTIR; Bruker ver. 113), to measure the extinction of the infrared radiation by the silicate dust particles suspended in an aerosol cloud filling the cell (diameter about 15~cm, length 0.6~m). By multiple reflection between the two gold mirrors at both ends of the cell, the IR radiation passed 32 times through the particle cloud, thus enhancing the sensitivity of the detection to optical depths around unity in the 10~$\mu$m band. An available dust flow generator (Palas RBG 1000), which utilized a rotating brush to disperse a pressed powder from a dust storage hole into a nitrogen gas stream, produced the dense aerosol that was led through stainless steel tubes toward the cell. In order to concentrate small-sized particles in the aerosol, we placed a two-stage impactor between the dust flow generator and the cell. The first stage retained clumps of particles above a certain size limit whereas the second stage densified the aerosol up to a concentration of 10$^6$ particles per cubic centimeter. A fraction of the aerosol was sampled through polyester membrane filters for scanning electron microscope (SEM) inspection of the particle morphology. After filling the aerosol into the cell, we stopped the aerosol flow from the generator and trapped the suspended particles in the cell during the spectroscopic analysis. Particles of 1~$\mu$m size and smaller are sufficiently coupled to the gas at atmospheric pressure to form a stable cloud on the timescale of hours. Larger particles and clumps of particles settled to the bottom of the cell. A small gas flow (50 liters hr$_{-1}$) from both ends of the cell toward the outlet valve was maintained during the measurement in order to protect the mirrors. This caused a continuous loss of aerosol and therefore a reduction of the particle column density during the experiment. Measurements shown here were usually taken after 10 minutes in order to allow for some sedimentation of remaining clumps. Apart from the decreasing absorption strength, there were no changes of the band profiles observed after this time. 

In order to make a comparative study, we have applied the classical KBr pellet technique where a granular sample (e.g. silicate particles) was mixed with potassium bromide powder (KBr), which has high transmission throughout the mid-infrared range, in a mass ratio of 1:500.  After very careful mixing in an agate mortar for segregating the particles as much as possible, 0.2~g of each mixture were pressed with 10~ton of force load to make about 0.5~mm thick clear pellets with a diameter of 1.3~mm, the transmission of which was measured in the FTIR spectrometer.

We investigated six powdered silicate samples of five different sorts with respect to composition and crystal structure. The crystalline Mg$_2$SiO$_4$ (forsterite) samples were commercial products whereas for the amorphous silicates we made use of original samples that were produced in our laboratory by the sol-gel technique \citep{jag03} and glass melting \citep{dor95}. Table~1 gives a list of the samples we used, together with their properties. We subjected the samples to a sedimentation process in a solvent (acetone), which allowed us to obtain a size fraction $leq$ 1.0~$\mu$m in diameter except CF2 and AF samples, which were originally small enough for the experiments. 
%
\begin{table*}
\caption{Properties of the silicate samples.}
\label{table:1}
\centering
\begin{tabular}{c c c c c c}
\hline
Sample & Chemical Formula & Product Information & Size ($\mu$m) & Shape \\  
\hline
CF1 & crystalline Mg$_2$SiO$_4$ &  Alfa Aesar Johnson & d$<$1.0       & irregular \\
CF2 & crystalline Mg$_2$SiO$_4$ &  Marusu Yuuyaku     & d$\approx$0.2 & ellipsoid\\
AF & amorphous Mg$_{2.3}$SiO$_4$ &  sol-gel process    & d$<$1.0        & irregular \\
CE & crystalline MgSiO$_3$     &  melting     & d$<$1.0        & irregular \\
AE & amorphous MgSiO$_3$     &  melting/quenching     & d$<$1.0        & irregular \\
CO & crystalline Mg$_{1.9}$Fe$_{0.1}$SiO$_4$ &  natural (San Carlos)   & d$<$1.0       & irregular \\
\hline
\end{tabular}
\end{table*}
%
\section{Results}
\subsection{Aerosol vs. KBr spectra}
The extinction spectra for all samples are shown in Figure 1 plotted as a normalized 
extinction efficiency versus the wavelength. The reason for normalizing 
all spectra to the maximum extinction values is that the aerosol measurements 
are not quantitative (see also $\S$~4). 

From Figure 1, it is obvious that for the crystalline silicates, 
the aerosol spectroscopy reveals considerably different band positions and 
also bandwidths when compared to the KBr measurements. The band positions and their 
differences are given as inserted tables in the plots. Especially for the strong 
bands of the olivine-type crystals (CF1, CF2, CO) at $\sim$9.8 and $\sim$11~$\mu$m, 
the spectra measured with the KBr pellet technique are shifted to longer wavelengths 
by amounts of up to 0.24~$\mu$m. Smaller features which do not shift very much, 
change their appearance from isolated bands into shoulders and vice versa due to 
the shift of the strong bands. For crystalline enstatite (CE), the shifts are not 
that strong (typically 0.1~$\mu$m) but affect all the band components. For the amorphous 
materials (AF and AE), there is a clear detectable shift with AE, but a distinct 
shift is not seen with AF. 

The fact that strong bands are more severely influenced than weaker ones is expected. 
The strong bands are known to be due to geometrical resonances 
of the grains, which, for example, for spherical grains occur at $\epsilon$\=-2$\epsilon_m$ 
where $\epsilon$ and $\epsilon_m$ are the dielectric functions of the silicate 
and the embedding medium (1.0 for air, 2.3 for KBr), respectively. This can probably 
explain the detected shift toward longer wavelengths, where $\epsilon$ 
decreases. For some theoretical simulations of band profiles for forsterite particles 
in vacuum and KBr, see \citet{fab01}. On the other hand, the aggregation state of the 
particles in the aerosol and the KBr is probably different because of the different dispersion 
methods and the pressing of the pellets. Therefore, it may cause an additional influence 
on the band profiles. Unfortunately, the agglomeration state in the KBr remains unknown, 
consequently the strength of this effect cannot be explored. 

Although in amorphous silicates negative values of $\epsilon$ and therefore geometrical 
resonances do not occur, a small band shift is expected for particles in KBr compared to 
aerosol measurements. Calculations  for the continuous distribution of ellipsoids (CDE) model and with Mie theory give 
$\Delta \lambda$~=~0.04 and 0.27~$\mu$m, respectively, for AE and 
$\Delta \lambda$~=~0.02 and 0.20~$\mu$m, respectively, for AF. It may have to do 
with the porous structure of the sol-gel material. Nevertheless, this shift 
is seen for AE but not for AF, which is currently not yet understood. 
%
\begin{figure*}
\figurenum{1}
\epsscale{0.92}
\plotone{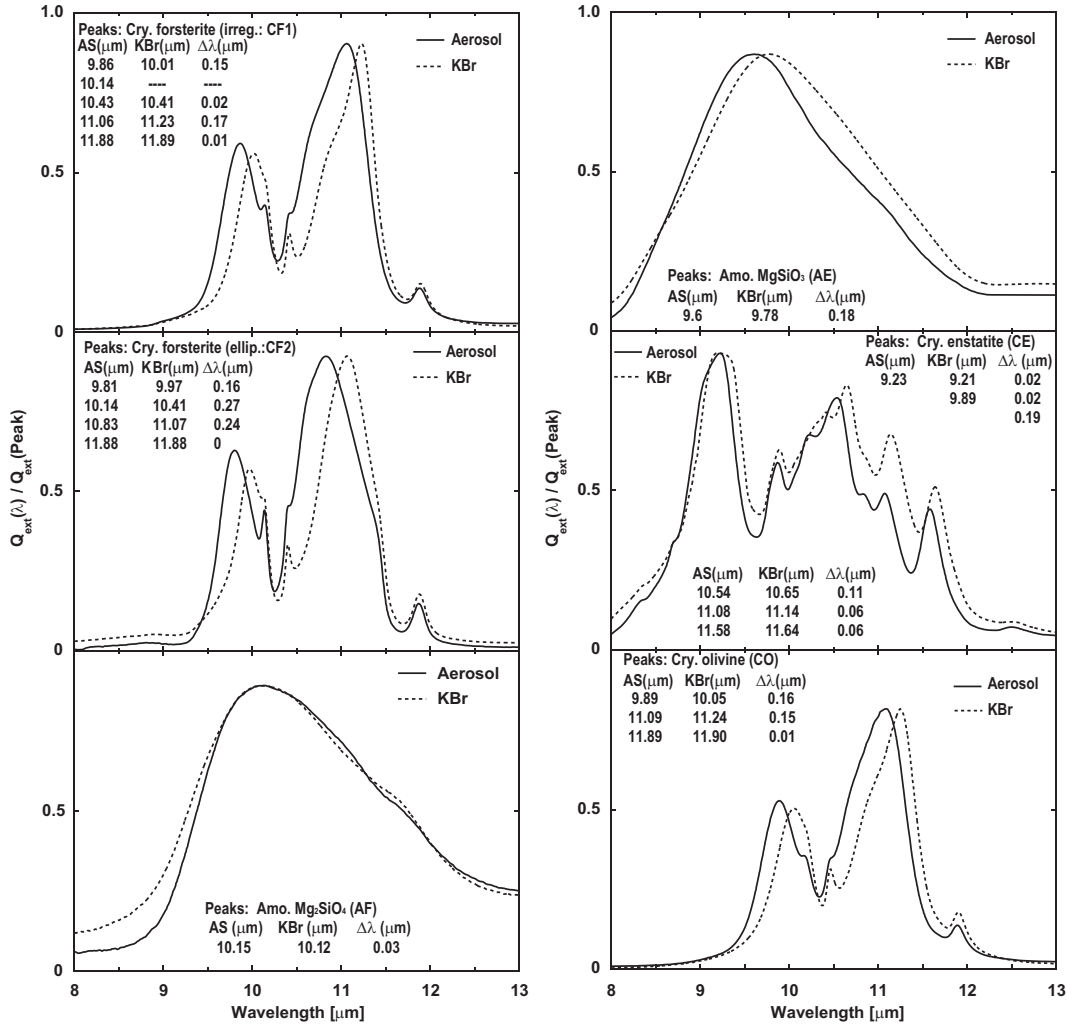}
\caption{Comparison of the Si-O stretching band profiles of {\it crystalline forsterite, 
amorphous {\rm Mg$_2$SiO$_4$}, crystalline enstatite, amorphous {\rm MgSiO$_3$}, and crystalline olivine} 
obtained from the aerosol (AS) and KBr measurements.}
\end{figure*}
\subsection{Effect of grain shape}
Figure 2 shows transmission electron microscope (TEM) and SEM images of CF1 and CF2 samples. The SEM pictures of the 
aerosol sampling demonstrate that most of the particles are in small and rather 
compact aggregates. The compactness indicates that they are clumps of the original 
powders and not agglomerates that have been grown in the aerosol. Although the 
clumps are less than 3~$\mu$m in size, the clumping should have an effect on the 
spectra. The close-packed clusters should tend to cause a broader band than the 
elongated ones as we have demonstrated by discrete dipole approximation (DDA) calculations for clusters of spherical 
SiO$_2$ particles \citep[for more details see][]{tam06}. 

Although it is not possible to obtain the exact aggregate morphology, as the 
TEM images are not in direct view of the aerosol sampling, more detailed grain shapes 
can be examined. The images clearly show that CF2 particles are rather roundish (ellipsoidal), 
whereas those of CF1 are irregular with sharp edges. The band profiles measured for 
these samples clearly differ in the sense that the peaks of the geometrical resonances 
are shifted by up to 0.23~$\mu$m toward shorter wavelengths for the roundish CF2 grains. 
Although it cannot be directly proven that this is a 
pure consequence of the grain shape, it is plausible because the geometrical resonances 
of a sphere are situated at such shorter wavelengths (compare Fig.~3). In contrast, 
irregular grains generally seem to produce geometrical resonances at longer wavelengths, 
i.e. closer to the transverse optical lattice frequencies. 
%
\begin{figure*}
\figurenum{2}
\epsscale{1.0}
\plotone{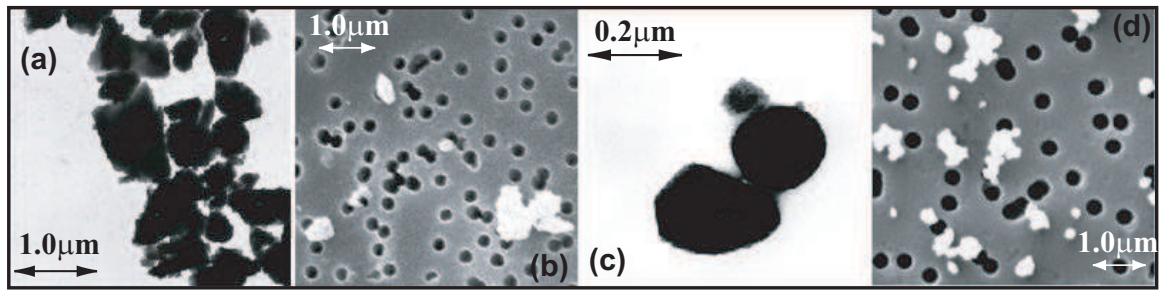}
\caption{(a,c) TEM and (b,d) SEM  and  images of the {\it crystalline forsterite} particles of samples (a,b) CF1 and (c,d) CF2. The SEM pictures show typical aggregates of 1-3~$\mu$m in size that have been directly sampled from the aerosol. 
The black dots in the SEM pictures are holes of the polyester filter having a diameter of 0.4~$\mu$m. The TEM 
pictures taken on the original powders give an impression of the grain shape. Note that sample CF2 has elliptical 
grains whereas those of sample CF1 are irregular. }
\end{figure*}
%
\subsection{Comparison with observations}
Finally, we demonstrate the importance of our new data by applying them to an observed spectrum. Figure 3 shows the emission spectrum of the crystalline silicate-rich Vega-type star HD113766 \citep{sch05}, together with the band profiles of CF1 measured via both the aerosol and the KBr measurements, 
as well as calculated band profiles of crystalline forsterite \citep[complex refractive index by][]{ser73} for the CDE model \citep{bh83}, and for the distribution of hollow spheres (DHS) model \citep[see][]{bo05}. \\
\begin{figure}
\figurenum{3}
\epsscale{1.2}
\plotone{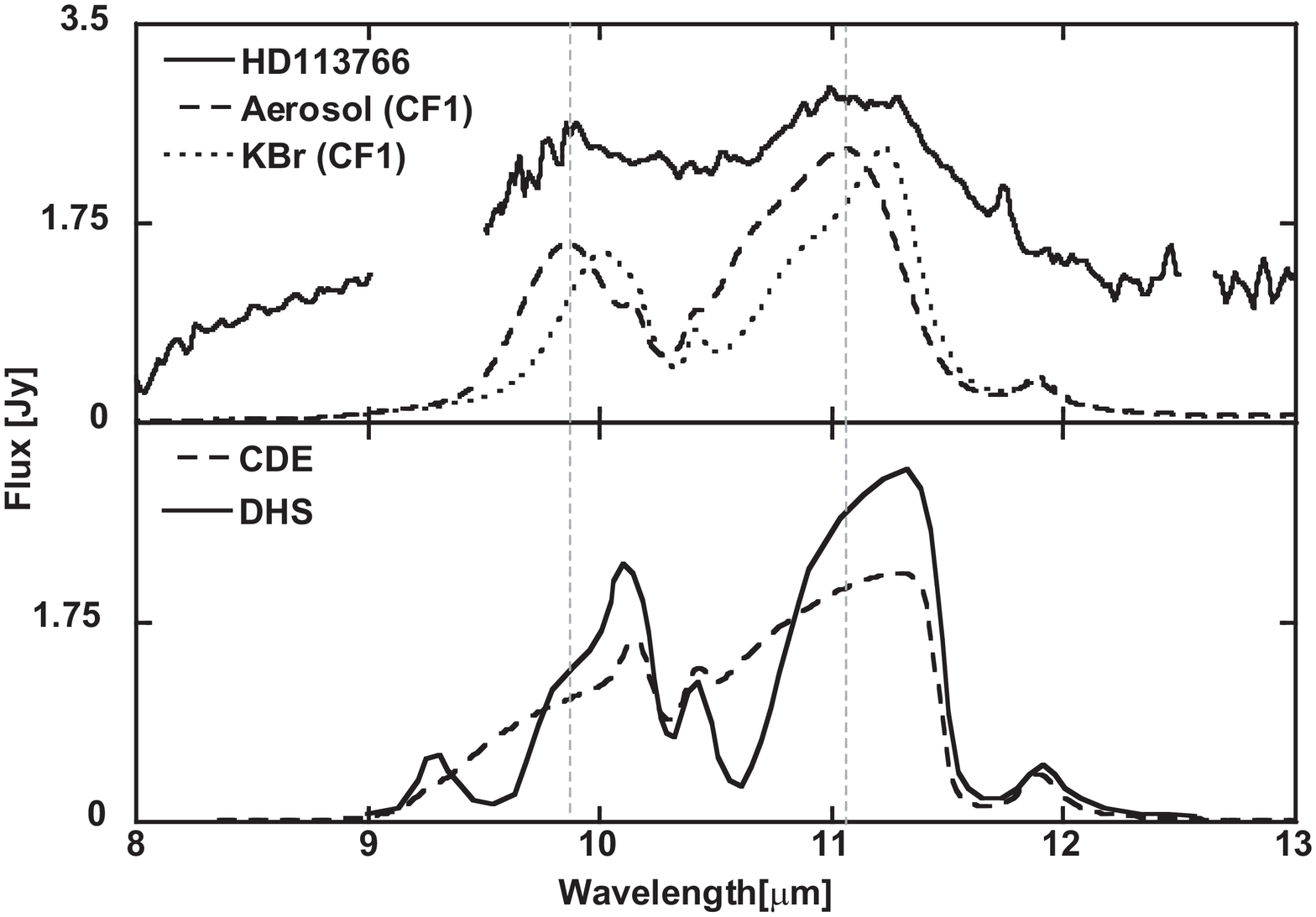}
\caption{{\it Top}: Comparison of the HD113766 emission feature with the CF1 band profile from the aerosol and KBr measurements. {\it Bottom}: Calculated band profiles for forsterite using the CDE and DHS models. 
The light dotted lines denote the peak positions of the aerosol result.}
\end{figure}
Evidently, the calculated band profiles cannot reproduce both the observed and the laboratory-measured 
data. The CDE model leads to band profiles that are far broader than the laboratory measurements. The DHS model 
predicts the structure of the observed band correctly but produces peaks at excessively long wavelengths, particularly at 10.1~$\mu$m (similar to the CDE). In contrast to this, the aerosol measurements of the 
CF1 sample reproduce quite well the positions of the main peaks in the observed spectrum at 9.9 and  11.0~$\mu$m. The peak at 11.3~$\mu$m, however, is not reproduced. This may indicate larger grains in the disk of the 
Vega-type star. We should note here that the measured spectrum of the natural olivine grains (sample CO) 
provides a similarly nice match whereas the band profile of sample CF2 fails to reproduce the observed 
features (see Fig.~1). This may indicate that the forsterite- or olivine-type grains in HD113766 differ 
in their morphological properties from the small roundish grains like CF2. 

\section{Conclusion: comparison of the aerosol and KBr techniques}
We have applied the aerosol technique to IR spectroscopic 
measurements of the 10~$\mu$m band profile of crystalline and amorphous 
silicates and demonstrated its application to obtaining realistic band 
profiles that can be compared directly to observed spectra for the first time. 

Compared to the traditional KBr technique, the extinction measurement in aerosol 
has two disadvantages: it requires about 100~mg of material and it is, unfortunately, 
not quantitative. The column density of the dust grains along the spectrometer 
beam can hardly be estimated or measured to an accuracy higher than a factor 
of 2. A better accuracy could be achieved only by accompanying quantitative 
KBr measurements or by a determination of the material's optical constants, 
for example, by reflection measurements. In such a case and given that a theoretical 
model can be adjusted to represent correctly the morphological properties of 
the particulate, the aerosol spectra can be renormalized. 

Conversely, there are distinct advantages of the aerosol method. First, it lacks 
the problem of the embedding medium, which is inherent to the KBr method. 
A second major advantage is that the structures 
of the grains can be analyzed by either in situ or ex situ microscopic imaging. 
This allows investigation in detail of the influence of the grain morphology on the 
spectra and will hopefully help to calibrate theoretical approaches. 

Concerning practical aspects, the preparation of the aerosol is easier than 
that of a KBr pellet with an ideal homogeneous grain distribution. Large 
clumps of particles can be receded or settled down within a short time. 
Even separation of different grain sizes is uncomplicated with the impactor technique. 
An important point is that collision timescales in the aerosol can probably be 
made smaller than the duration of an experiment, so that fractal aggregate formation 
could be directly studied by IR spectroscopy. So far, the particle densities were 
too small in our experiment to observe grain growth by aggregation. 
With the KBr technique, this is hardly possible, because agglomerates are 
modified or destroyed by the preparation process. Finally, with the aerosol technique, 
it should be possible to manipulate the dust in situ by surface condensation or irradiation, 
so that its influences on the IR spectra can be measured well.
\acknowledgements
Our project has been supported by Deutsche Forschungsgemeinschaft (DFG) under the grant MU 1164/5-3/4. 
We express our gratitude to C.~Koike for providing us the Marusu (CF2) sample and C.~J\"ager 
for supplying the enstatite  and sol-gel samples. We are grateful to O.~Sch\"utz for sharing his 
observational data. We appreciate great technical supports by W.~Teuschel, G.~Born, and K.~Tachihara.

\end{document}